\documentclass{aa}
\usepackage{amsmath}
\usepackage[varg]{txfonts}
\usepackage{graphicx}
\usepackage{epstopdf}
\usepackage{natbib}
\usepackage{float}
\usepackage{lscape}
\usepackage{longtable}

\begin{document}

\title{The effects of stellar winds on the magnetospheres and potential habitability of exoplanets}

\author{V. See\inst{1}
\and M. Jardine\inst{1}
\and A. A. Vidotto\inst{1,2}
\and P. Petit\inst{3,4}
\and S. C. Marsden\inst{5}
\and S. V. Jeffers\inst{6}
\and J.D. do Nascimento Jr.\inst{7,8}}

\institute{SUPA, School of Physics and Astronomy, University of St Andrews, North Haugh, KY16 9SS, St Andrews, UK \\ \email{wcvs@st-andrews.ac.uk}
\and Observatoire de Gen\`eve, Universit\'e de Gen\`eve, Chemin des Maillettes 51, Sauverny, CH-1290, Switzerland
\and Universit\'{e} de Toulouse, UPS-OMP, Institut de Recherche en Astrophysique et Plan\'{e}tologie, Toulouse, France
\and CNRS, Institut de Recherche en Astrophysique et Plan\'{e}tologie, 14 Avenue Edouard Belin, F-31400 Toulouse, France
\and Computational Engineering and Science Research Centre, University of Southern Queensland, Toowoomba, 4350, Australia
\and Universit\"at G\"ottingen, Institut f\"ur Astrophysik, Friedrich-Hund-Platz 1, 37077 G\"ottingen, Germany
\and Departmento de F\'isica Te\'orica e Experimental, Universidade Federal do Rio Grande do Norte, CEP:59072-970 Natal, RN, Brazil
\and Harvard-Smithsonian Center for Astrophysics, Cambridge, Massachusetts 02138, USA}

\abstract {The principle definition of habitability for exoplanets is whether they can sustain liquid water on their surfaces, i.e. that they orbit within the habitable zone. However, the planet's magnetosphere should also be considered, since without it, an exoplanet's atmosphere may be eroded away by stellar winds.} {The aim of this paper is to investigate magnetospheric protection of a planet from the effects of stellar winds from solar-mass stars. } {We study hypothetical Earth-like exoplanets orbiting in the host star's habitable zone for a sample of 124 solar-mass stars. These are targets that have been observed by the Bcool collaboration. Using two wind models, we calculate the magnetospheric extent of each exoplanet. These wind models are computationally inexpensive and allow the community to quickly estimate the magnetospheric size of magnetised Earth-analogues orbiting cool stars.} {Most of the simulated planets in our sample can maintain a magnetosphere of $\sim$5 Earth radii or larger. This suggests that magnetised Earth analogues in the habitable zones of solar analogues are able to protect their atmospheres and is in contrast to planets around young active M dwarfs. In general, we find that Earth-analogues around solar-type stars, of age 1.5 Gyr or older, can maintain at least a Paleoarchean Earth sized magnetosphere. Our results indicate that planets around 0.6 - 0.8 solar-mass stars on the low activity side of the Vaughan-Preston gap are the optimum observing targets for habitable Earth analogues.} {}


\maketitle

\section{Introduction}
The search for exoplanets has been going on for nearly two decades \citep{Mayor1995,Marcy1996,Butler1996} with an emphasis on finding habitable extra-solar Earth analogues \citep{Kaltenegger2009,Fressin2012,Borucki2012}. A commonly used criterion when assessing the suitability of a planet for life is whether it lies within the so-called habitable zone (henceforth HZ). This region is defined as the set of orbital distances at which liquid water can exist on the planetary surface \citep{Huang1960,Hart1978,Kasting1993,Kopparapu2013} and is determined by considering the flux of radiation incident on the planet.

It is also important to consider other factors. A stable atmosphere, which is necessary to regulate surface temperatures, could be eroded away by sufficiently strong stellar winds \citep{Khodachenko2007,Zendejas2010,Vidotto2011,Lammer2012} rendering a planet uninhabitable. Earth has retained its atmosphere thanks to the shielding provided by its magnetosphere. In contrast, Mars and Venus both lack a substantial intrinsic magnetic field. As a result, both suffer significant atmospheric losses with Mars having a much thinner atmosphere \citep{Wood2006,Edberg2010,Edberg2011}. 

We therefore have at least two distinct parameters that need to be considered when searching for extra-solar Earth analogues. On the one hand, the exoplanet should lie within the habitable zone allowing for Earth-like surface temperatures. On the other hand, the exoplanet should have a magnetosphere large enough to shield the atmosphere.

In this paper, we assess the ability of exoplanets, around solar-type stars, to maintain magnetospheres similar in size to both the present day and early Earth's magnetospheres. The thermal plasma pressure, wind ram pressure, and stellar magnetic pressure all act to compress the exoplanetary magnetosphere. \citet{Vidotto2013} have studied how the stellar magnetic pressure affects hypothetical Earth analogues around M dwarfs. Compared to solar-type stars, M dwarfs have close-in HZs and can possess much stronger magnetic field strengths \citep{Donati2008,Morin2008,Morin2010}. The stellar magnetic pressure is therefore the dominant pressure term of the wind in the HZ. In contrast, the ram pressure dominates in the HZ of solar-type stars \citep{Zarka2001,Zarka2007,Jardine2008} and so we limit ourselves to studying the stellar wind ram pressure term in this paper.

In order to study the interaction between stellar winds and exoplanets, we use a survey of 167 stars observed by the Bcool collaboration \citep{Marsden2013}. We excluded the subgiants and any stars that did not have all the data required for our wind models. This left 124 solar-type stars, mostly with masses between 0.8$M_{\odot}$ and 1.4$M_{\odot}$. We refer the reader to \citet{Marsden2013} for full details of the sample. We will assume the existence of a fictitious exoplanet orbiting in the HZ of each star in our sample. Unfortunately, typical exoplanetary magnetic field strengths are not known since there have been no direct observations of exoplanetary magnetic fields to date, although \citet{Vidotto2010} and \citet{Llama2011} hint at a possible indirect detection. In light of this, we assume that the fictitious planets have the same properties as Earth, i.e. same mass, radius, and magnetic field strength. For each Earth analogue, we calculate the ram pressure exerted on it and determine if it can maintain a present day Earth-sized magnetosphere. 

\citet{Lammer2007} suggest that smaller magnetospheres may still offer adequate protection and it is thought the Earth had a smaller magnetosphere in its past as a result of higher solar activity \citep{Sterenborg2011}. \citet{Tarduno2010} report that the Earth had a geodynamo around 3.4 Gyr ago, during the Paleoarchean, which generated a magnetic field that was roughly 50\% weaker than the present day's. Using this field strength and the empirical wind model of \citet{Wood2006}, \citet{Tarduno2010} estimate a magnetosphere size of around 5 $R_E$. Since the Earth was able to retain its atmosphere, it is reasonable to assume that a Paleoarchean sized magnetosphere would sufficiently protect an Earth analogue. However, the magnetospheric size estimate of \citet{Tarduno2010} is dependent on the wind model adopted. We discuss the range of possible Paleoarchean magnetosphere sizes using different models in our results.

The rest of the paper is structured as follows. Section \ref{secStellarWindModel} covers the details of the wind models used. Section \ref{secResults} covers the results obtained using the models outlined in the previous section and their broader implications within the context of other works. A discussion and concluding remarks follow in Sect. \ref{secConclusion}.

\section{Stellar wind models}
\label{secStellarWindModel}
Modelling stellar winds is a difficult task since their acceleration mechanisms are still uncertain. Early models simply admitted free parameters in the form of an isothermal wind temperature or a polytropic index with values chosen to match solar wind observations \citep{Parker1958,WeberDavis1967,Sakurai1985,KeppensGoedbloed1998}. This is not feasible for stellar winds because of the lack of in-situ measurements, although indirect detections do exist \citep{Wood2005}. Other attempts have been made to characterise the mass loss rates from stars through the use of semi-empirical scaling relations, notably by \citet{Reimers1975, Reimers1977} and subsequently \citet{SchroderCuntz2005}. However, these approaches still contain a fitting parameter that require observations to calibrate. 

The most up to date models are fully three dimensional and self-consistent \citep{Vidotto2009,Vidotto2012} and use more physically motivated arguments to determine mass loss rates \citep{HolzwarthJardine2007,Cohen2007,Cranmer2011}. However many assumptions are used and scaling relations are still present in some models. While no current consensus exists on which mechanism is responsible for heating stellar coronae, two front runners have emerged - wave driven heating \citep{Cranmer2007,vanB2011} and reconnection driven heating \citep{FiskZhao2009}. The complexity of these models indicates the level of difficulty in accurately determining the mass loss rates or ram pressures.

For this work, we employ the Parker wind model \citep{Parker1958} and the Cranmer \& Saar (henceforth CS) model of mass loss \citep{Cranmer2011} which are both one-dimensional models. We favour these simpler models over the more sophisticated alternatives mentioned previously for several reasons. Chiefly, it would take a prohibitively long time to process a sample of this size with a full MHD code. Additionally, the full set of input parameters that these models require, e.g. magnetic maps of the stellar surface, does not exist for more than a few stars in this sample. An additional benefit is the ease with which these models can be implemented. This equips the community with a tool to quickly determine the magnetospheric size of any given exoplanet in future habitability studies.

\subsection{Parker wind model}
\label{subSecParkerModel}

\begin{table*}
	\caption{The numerical values used in this study for the present day Earth's magnetic moment \citep{Tarduno2010}, solar coronal number density \citep{Guhathakurta1996}, luminosity \citep{Harmanec2011}, X-ray luminosity \citep{Judge2003}, chromospheric activity \citep{Mamajek2008}, and mass loss rate \citep{Cranmer2011}. In addition, the values for the velocity (chosen to be roughly the escape velocity), number density \citep{Balikhin1993}, and ram pressure ($P_{ram}^E=mn^E_{sw} (v^E_{sw})^2$) of the solar wind in the vicinity of the Earth are also listed.}
	\centering
	\begin{tabular}{ccccccccc}
		\hline\hline
		$M_{E}$ & $\bar{n}_{c\odot}$ & $L_{\odot}$ & $L_{X\odot}$ & $\log R'_{HK\odot}$ & $\dot{M}_{\odot}$ & $v^{E}_{sw}$ & $n^{E}_{sw}$ & $P^{E}_{ram}$\\
		$[\mathrm{Am}^{2}]$ & $[\mathrm{cm}^{-3}]$ & $[\mathrm{erg} \mathrm{s}^{-1}]$ & $[\mathrm{erg} \mathrm{s}^{-1}]$ & $[\mathrm{dex}]$ & $[\mathrm{M}_{\odot}\mathrm{yr}^{-1}]$ & $[\mathrm{km s}^{-1}]$ & $[\mathrm{cm}^{-3}]$ & $[\mathrm{nPa}]$\\
		\hline
		$8 \times 10^{22}$ & $10^8$ & $3.85 \times 10^{33}$ & $2.24 \times 10^{27}$ & $-4.905$ & $1.90 \times 10^{-14}$ & $600$ & $5$ & $1.8$\\
		\hline	
	\end{tabular}
	\label{tabParams}
\end{table*}

The Parker wind model \citep{Parker1958} assumes a steady, isothermal, and spherically symmetric wind. The velocity is given by integrating the momentum equation,

\begin{equation}
	\rho m v\frac{\partial v}{\partial r}=-\frac{\partial}{\partial r}(\rho k_BT)-\rho m \frac{GM_{\star}}{r^2},
	\label{eqParkerWind}
\end{equation}
where $\rho$, $v$, and $T$ are the mass density, velocity and temperature respectively, $r$ is the radial distance from the stellar object, $m$ is the mean particle mass, taken to be 0.6 times the proton mass, and $M_{\star}$ is the stellar mass. We look for flows that are subsonic at small radii and supersonic at large radii with the transition between the two regimes occurring at the sonic point, $r_{s}=GM_{\star}/2c_s^2$ where $c_s$ is the isothermal sound speed.

The density profile is then obtained from mass conservation,

\begin{equation}
	\tilde{\rho}=\frac{\rho}{\rho_{0}}=\frac{v_{0}r_{0}^{2}}{vr^{2}},
	\label{eqMassConservation}
\end{equation}
where $\tilde{\rho}$ is the density normalised to $\rho_{0}$, the density at an arbitrary position, $r_{0}$. The velocity has a corresponding value $v_{0}$ at $r_{0}$. We choose $r_{0}$ to be the stellar radius, i.e. $r_{0}=r_{\star}$. The density at the stellar surface can be estimated from the chromospheric activity using a relation for the X-ray luminosity, $L_X=\Lambda\left(T\right) EM$, where EM is the emission measure and $\Lambda (T)$ is the radiative loss function, and an empirically obtained relation between stellar chromospheric activity and X-ray luminosity \citep{Mamajek2008},

\begin{equation}
	\log{R_{X}}=(-4.90 \pm 0.04)+(3.46 \pm 0.18)(\log{R'_{HK}}+4.53).
	\label{eqActivity}
\end{equation}
Here, $R_{X} = L_{X}/L_{bol}$, is the X-ray luminosity normalised to the bolometric luminosity and $R'_{HK}=L_{HK}/L_{bol}$ is the chromospheric activity measured in Ca II H \& K lines. The emission measure is a quantity that characterises the amount of free-free emission originating from a volume of electrons and is given by

\begin{equation}
	EM=\int n^2_{ec} dV \simeq \bar{n}^{2}_{ec} \cdot \frac{4}{3}\pi r^3_{\star} \left(\frac{r^3_{ss}}{r^3_{\star}}-1\right),
	\label{eqEmissionMeasure}
\end{equation}
where $n_{ec}$ is the closed coronal electron number density and $\bar{n}_{ec}$ is the coronal electron number density averaged over the emitting region. This region is assumed to be a shell of uniform density extending from the stellar surface, $r_{\star}$, to an outer surface known as the source surface, $r_{ss}$ \citep{AltschulerNewkirk1969,Jardine2002}. The source surface represents the limit of confinement of the corona, beyond which the pressure of the hot coronal gas opens up the magnetic field lines. Since the HZ of solar-type stars lies far beyond the source surface, we may ignore the magnetic pressure of the star. Normalised to solar values, we therefore have

\begin{equation}
	\frac{L_{X\star}}{L_{X\odot}}=\left(\frac{\bar{n}_{e c \star}}{\bar{n}_{e c \odot}}\right)^{2}\left(\frac{r_{\star}}{r_{\odot}}\right)^3=\left(\frac{\rho_{c \star}}{\rho_{c \odot}}\right)^{2}\left(\frac{r_{\star}}{r_{\odot}}\right)^3,
	\label{eqXrayRatio}
\end{equation}
where $\rho_{c\odot}=m\bar{n}_{c\odot}$. We have assumed that the radiative loss function will take on similar values over the range of temperatures present in our sample of stars and that $r_{ss}/r_{\star}$ is roughly constant. Quasi-neutrality has been assumed to obtain the final equality. The subscripts $\star$ and $\odot$ indicate stellar and solar parameters respectively and the subscript $c$ attached to the densities in Eq. (\ref{eqXrayRatio}) indicate these are densities in the closed corona. Substituting Eq. (\ref{eqXrayRatio}) into Eq. (\ref{eqActivity}) for $L_{X\star}$ and re-arranging, we obtain an estimate of the closed coronal density. Omitting the errors from Eq. (\ref{eqActivity}) for clarity, this is given by

\begin{equation}
	\rho_{c\star}=\rho_{c\odot}\left(\frac{r_{\odot}^3}{r_{\star}^3}\frac{L_{bol\star}}{L_{X\odot}}10^{-4.90+3.46(\log{R'_{HK}}+4.53)}\right)^{\frac{1}{2}}.
	\label{eqBaseDensity}
\end{equation}
On the Sun, the densities of coronal holes and the closed corona are known to differ. For the density at the base of the wind, we scale the coronal density calculated in Eq. (\ref{eqBaseDensity}) by a factor, $f$, and adopt a value of $f=0.1$ \citep{Guhathakurta1996}. The ram pressure of the stellar wind at a given radial distance from the host star can therefore be determined by Eqs. (\ref{eqParkerWind}), (\ref{eqMassConservation}), \& (\ref{eqBaseDensity}) and is given by

\begin{equation}
	P_{ram}(r)=\rho(r) v^{2}(r)=\rho_{c\star} f \tilde{\rho}(r)v^{2}(r).
	\label{eqRamPressure}
\end{equation}
The only variable in Eq. (\ref{eqRamPressure}) that is unconstrained by the model we have presented is the wind temperature. We adopt a temperature of 2.1MK since this reproduces the solar wind parameters at Earth to a good degree within our model: $v=1.15$ $v_{sw}^{E}$, $n=0.74$ $n_{sw}^{E}$. Table \ref{tabParams} contains the numerical values for various parameters used in this study.

\subsection{Cranmer \& Saar model}
\label{subsecCranmerWind}
\citet{Cranmer2011} calculate mass-loss rates by considering the basal Alfv\'en wave energy flux emerging through the photosphere. The Alfv\'en waves either turbulently heat the corona sufficiently for a gas pressure wind to be driven or drive a wind directly by wave action. The total mass-loss rate is calculated with contributions from both mechanisms.

The advantage of this model is that all the parameters required (stellar mass, radius, luminosity, metallicity, and rotation period) can be observationally obtained. The first four are determined in the Bcool sample but rotation periods are unknown. In order to estimate the rotation periods, we determine the Rossby number, $R_0$, from the chromospheric activity relation of \citep{Mamajek2008} which is given by

\begin{equation}
\begin{array}{cc}
R_{0}=(0.808 \pm 0.014)-(2.966 \pm 0.098)\left(\log R'_{HK}+4.52\right)\ \\
R_{0}=(0.233 \pm 0.015)-(0.689 \pm 0.063)\left(\log R'_{HK}+4.23\right).\
\end{array}
\label{eqRossbyActivity}
\end{equation}
The upper equation applies for $-5 \leq \log R'_{HK} < -4.3$ and the lower equation for $\log R'_{HK} \geq -4.3$. The Rossby number is the stellar rotation period normalised to the convective turnover time and has been shown to be an effective probe into stellar activity  \citep{Pizzolato2003,Wright2011}. For each star we then iterate with the CS model until we find the rotation period that is consistent with this Rossby number. As a result of the presence of low activity stars in the Bcool sample, we extend the upper relation into the $\log R'_{HK} < -5$ regime. Although it may be advisable to treat the $\log R'_{HK} < -5$ results with slightly more skepticism, we note that they do not differ greatly from the results obtained using the Parker model for the stellar wind. Ram pressures can then be calculated using

\begin{equation}
	P_{ram}\left(r\right) = \frac{\dot{M}v_{esc}}{4\pi r^2},
	\label{eqCranPram}
\end{equation}
where $\dot{M}$ is the mass loss rate calculated using the Cranmer \& Saar model and $v_{esc}$ is the escape velocity of the star which the model uses as an estimate of the terminal wind velocity.

\begin{figure*}
	\begin{center}
	\includegraphics[width=\textwidth]{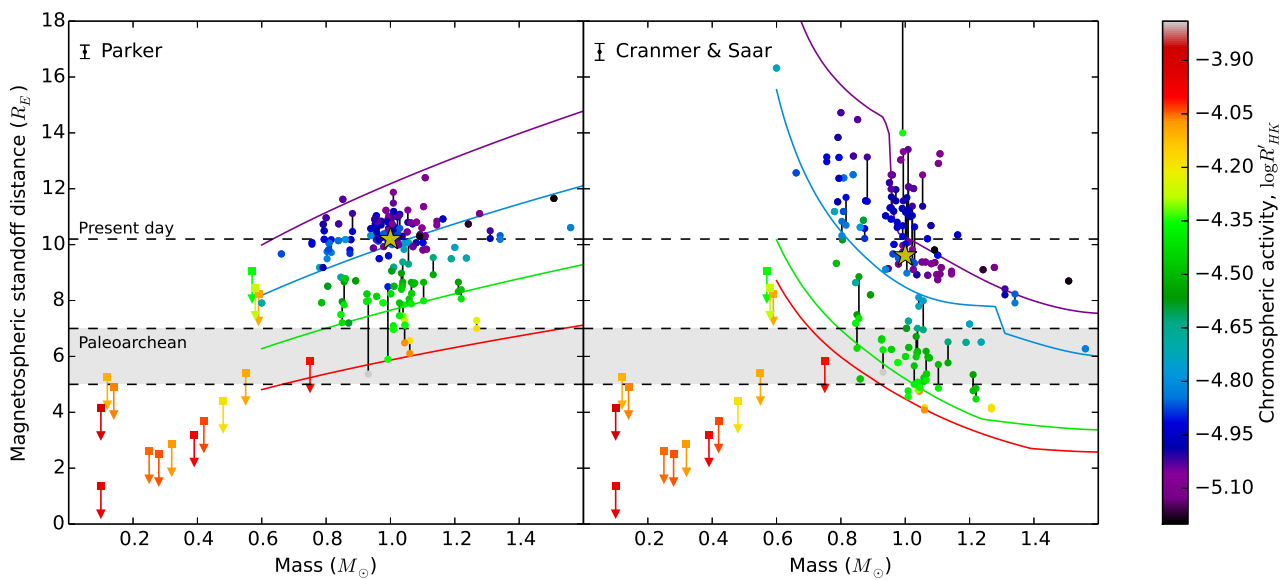}
	\end{center}
	\caption{Magnetospheric size as a function of host star mass for the Parker (left) and CS (right) wind models. The Bcool sample of solar-type stars is plotted with circles and the Sun is indicated by a star symbol. Values for these standoff distances can be found in App. A. Typical errorbars for this sample are indicated in the upper left of each panel. The magnetospheric sizes computed assuming constant activity, $\log R'_{HK}=-5.1,-4.8,-4.4,-4.0$, are plotted as solid lines. These correspond to chromospheric ages of 8.4, 3.2, 0.3, 0.008 Gyr respectively. We note that the Earth developed an oxygen rich atmosphere near 1.5 Gyr which corresponds to a solar chromospheric activity of $\log R'_{HK}=-4.6$. The upper limit for magnetospheric sizes as calculated by \citet{Vidotto2013}, for a sample of M dwarfs, are plotted with squares.  All points and curves are colour coded by chromospheric activity. The upper dashed line indicates the present day magnetosphere size and the shaded area indicates a range of possible Paleoarchean magnetosphere sizes.}
	\label{figMPHZ}
\end{figure*}

\section{Results and discussion}
\label{secResults}
\subsection{Magnetospheric extent within the habitable zone}
\label{subsecPRamHZ}
If the ram pressure acting against an Earth-like magnetosphere is known, then its size can be calculated by balancing the wind pressure with the planetary magnetospheric pressure \citep{Griessmeier2004}, i.e.

\begin{equation}
	r_{MP} = \left(\frac{\mu_0 f^2_0 M^2_E}{8\pi^2 P_{ram}}\right)^{1/6},
	\label{eqRMP}
\end{equation}
where $f_0$, taken to be 1.16, is a form factor included to account for the non-spherical shape of Earth's magnetosphere, and $M_E$ is the Earth's magnetic moment, taken to be 8 $\times$ 10$^{22}$Am$^2$. In Fig. \ref{figMPHZ} we show the magnetospheric size of each planet when located at the centre of the HZ (circular points), considering the stellar wind model given by Parker (left panel) and by Cranmer \& Saar (right panel). The HZ boundaries are calculated using the formulation of \citet{Kasting1993} with the water loss and first condensation limits defining the inner and outer edges respectively. Typical error bars for the magnetospheric sizes, calculated by propagating through the errors in Eqs. (\ref{eqActivity}) and (\ref{eqRossbyActivity}), are shown in the upper left corner. They are relatively small and we simply note that the scatter in Eqs. (\ref{eqActivity}) and (\ref{eqRossbyActivity}) has little impact on our conclusions. Additionally, stars with a large range of activities have upper and lower limits connected by a line. The numerical values of the five fictitious planets with the largest magnetospheric sizes, in both models, can be found in Table \ref{tabCalcVars}. Magnetosphere sizes, as well as the ram pressures at the planets, mass-loss rates and HZ locations, for the rest of the sample are available in App. A. We have also plotted several theoretical magnetospheric standoff curves as a function of stellar mass, each at a constant chromospheric activity. These are numerically calculated using the models presented in Sect. \ref{secStellarWindModel}, assuming the scaling relations $r_{\star} \propto M_{\star}^{0.8}$ and $L_{\star} \propto M_{\star}^{4}$. Metallicities are all set to solar levels in the CS model. We limit these curves to the stellar masses present in the Bcool sample. Both the data points and curves are colour coded by host star chromospheric activity.

\begin{figure*}
	\begin{center}
	\includegraphics[width=\textwidth]{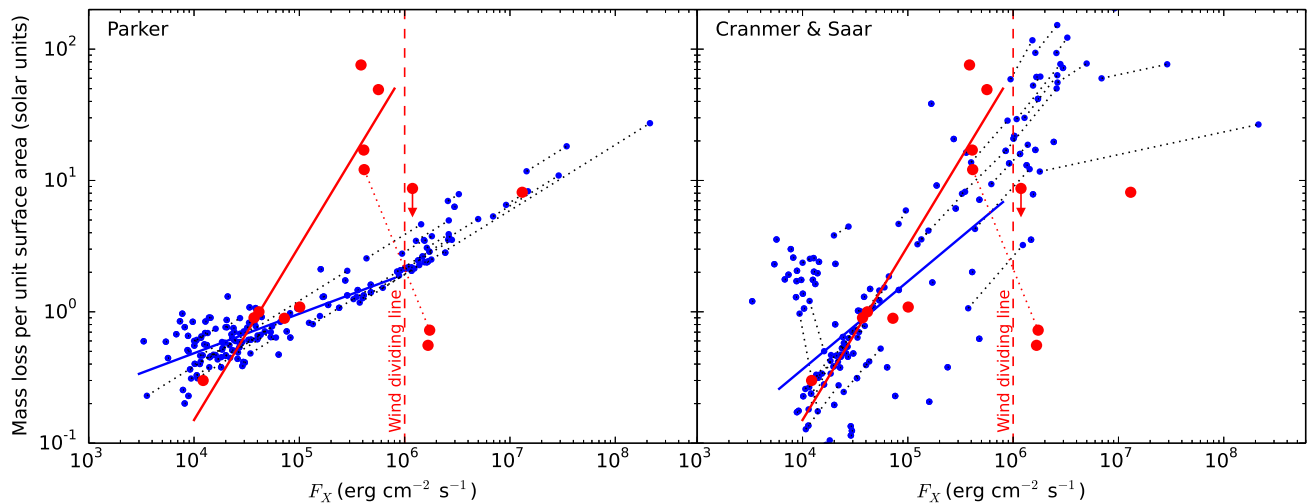}
	\end{center}
	\caption{Mass-loss rates per unit surface area as a function of X-ray flux for the Parker (left) and CS (right) models are plotted in blue. Values for these mass-loss rates can be found in App. A. The mass-loss rates and the wind dividing line of \citet{Wood2014} are overplotted in red (see their Fig. 4). These authors fit a power law (red solid line), $\dot{M} \propto F_X^{1.34\pm0.18}$, to their data points below the wind dividing line. For our data, we find power laws of $\dot{M} \propto F_X^{0.30\pm0.02}$ and $\dot{M} \propto F_X^{0.67\pm0.10}$ for the Parker and CS models respectively (solid blue lines). If we use our entire sample, we find $\dot{M} \propto F_X^{0.38\pm0.01}$ and $\dot{M} \propto F_X^{0.79\pm0.05}$ respectively. We note that our quoted errors are in the fit only.}
	\label{figWood}
\end{figure*}

In both models, the magnetospheric size increases with decreasing activity. The data points follow the curves, albeit with some scatter due to departures from the radius and luminosity scaling relations. For our modest sample of stars, a fraction of the planets can maintain present-day sized magnetospheres for both models. Many more of the planets can maintain a Paleoarchean sized magnetosphere. Using the wind model of \citet{Wood2006} and a reduced terrestrial magnetic moment of $4.8 \times 10^{22}$ Am$^2$, \citet{Tarduno2010} estimate the Earth's magnetosphere size to be 5 $R_E$ during the Paleoarchean when the Sun was 1.2 Gyr old. At this age, the Sun would have had a chromospheric activity of -4.6 according to the activity-age relation of \citet{Mamajek2008}. Using this activity level and the reduced magnetic moment, we estimate the Paleoarchean terrestrial magnetosphere size to be 7.0 $R_E$ and 5.3 $R_E$ for the Parker and CS models respectively. We therefore take 5 $R_E$ to 7 $R_E$ to be a plausible range of values for the magnetosphere size during the Paleoarchean. While most of the planets have magnetospheres larger than this range of Paleoarchean magnetosphere sizes in the Parker model, a significant number of them, in the CS model, fall in or below it. The level of chromospheric activity is therefore important in determining if a magnetosphere larger than that of the young Earth can be sustained. Additionally, this highlights the impact that a different wind model can have.

Stellar activity is believed to be a function of stellar age \citep{Skumanich1972,Soderblom1991,Donahue1998,Mamajek2008,Vidotto2014}. As stars spin down with age, their chromospheric activity, as well as magnetic activity in general, falls, resulting in larger planetary magnetospheres. Using the chromospheric age-activity relation given by \citet{Mamajek2008}, it is possible to gain a sense of the time evolution over the Bcool sample. We note that chromospheric ages are only indicative of the true age and that stellar ages are, in general, difficult to determine. Most stars have activities between -4.4 (green curve) and -5.1 (purple curve) which correspond to ages of 0.3 Gyr and 8.4 Gyr respectively. In order to place this in context, we consider an age of 1.5 Gyr, i.e. 3 Gyr ago, close to the age at which the Earth developed an oxygen-rich atmosphere \citep{Crowe2013}. This corresponds to a chromospheric activity of -4.6 which puts it roughly midway between the blue and green curves. At this age, the CS model suggests that Earth-like planets should have a minimum magnetospheric size of 5 $R_E$ in the mass range 0.6 $M_{\odot}$ - 1.6 $M_{\odot}$. Standoff distances are larger for the Parker model at this age with a minimum size of 7 $R_E$ in the same mass range. Both models agree that stars do not have to be very old before any Earth-like planets, that they are hosting, are able to maintain, at least, a 5 $R_E$ magnetosphere. For the CS model, this is around 1.5 Gyr and for the Parker model, it is likely to be almost immediately after the star enters the main sequence. In order for a planet orbiting in the HZ of a 1.5 Gyr old star to maintain a 7 $R_E$ magnetosphere under the CS model, the host star needs to be less than 1.0 $M_{\odot}$.

\begin{table}
\caption{Magnetospheric standoff distances, $r_{MP}$ for the five fictitious planets with the largest values of $r_{MP}$ in the Parker (top 5 results) and CS (bottom 5 results) models. Superscripts indicate the model used. Stars with large activity ranges have minimum and maximum magnetospheric sizes listed. Ram pressure and magnetospheric sizes for the entire sample are shown in Table \ref{tabCalcVarsApp}.}
\label{tabCalcVars}
\centering
\begin{tabular}{lcccc}
\hline\hline
Star & $r_{MP}^{P}$ & $r_{MP}^{CS}$ \\
ID & [$R_E$] & [$R_E$]\\
\hline
HD 217107	&	12.39	&	13.25	\\
HD 98618	&	11.87/11.49	&	10.08/13.41	\\
HD 107213	&	11.65	&	8.7	\\
HD 3765	&	11.62	&	14.48	\\
HD 28005	&	11.36	&	12.91	\\
\hline					
$\xi$ Boo B	&	8.50/5.90	&	31.76/14.00	\\
HD 88230	&	7.91	&	16.32	\\
HD 122064	&	10.96	&	14.72	\\
HD 3765	&	11.62	&	14.48	\\
HD 166620	&	10.73	&	13.84	\\
\hline	
\end{tabular}
\end{table}

The main difference between the two models presented in this paper is the mass dependent behaviour. The CS model shows decreasing standoff distances with mass which is the opposite behaviour in the Parker model. It is always possible to reconcile the two models by adjusting the wind temperature adopted in the Parker model. More massive, and hence hotter, stars have higher convective velocities resulting in a higher wave energy flux emerging through the photosphere \citep{Musielak2002}. More energy is therefore available to drive the wind within the CS model. This increased wave energy flux would lead to higher coronal temperatures and could be mimicked in a Parker-type model by assuming a mass-dependent coronal temperature. However, most of the stars in our sample do not have observational constraints of their wind properties (e.g. using the technique of \citet{Wood2006}). As a result, it is difficult to select which wind model (Parker or CS) is the most appropriate one.

\subsection{Comparison with other work}
\label{subsecOtherWork}
We compare the mass-loss rates that result from our models with ones obtained indirectly from observations of astrospheric Ly$\alpha$ absorption (see \citet{Wood2014} and references therein). Of their stars, three are also present in our own sample: $\epsilon$ Eri with a mass-loss rate of $30\dot{M}_{\odot}$ \citep{Wood2002} and the $\xi$ Boo binary system with a combined mass-loss rate of $5\dot{M}_{\odot}$ \citep{Wood2005}. For $\epsilon$ Eri, we obtain mass-loss rates of $0.73\dot{M}_{\odot}$ - $1.33\dot{M}_{\odot}$ (Parker) and $0.66\dot{M}_{\odot}$ - $2.01\dot{M}_{\odot}$ (CS). For $\xi$ Boo A, we obtain mass-loss rates of $1.86\dot{M}_{\odot}$ - $20.3\dot{M}_{\odot}$ (Parker) and $8.69\dot{M}_{\odot}$ - $19.8\dot{M}_{\odot}$ (CS) while for $\xi$ Boo B we obtain $0.28\dot{M}_{\odot}$ - $2.47\dot{M}_{\odot}$ (Parker) and $9.7\times10^{-5}\dot{M}_{\odot}$-$0.013\dot{M}_{\odot}$ (CS). Although the estimates do not agree with observations, this is not too much of a concern for the present work because of the 1/6 power dependence in Eq. (\ref{eqRMP}). The $\xi$ Boo estimates are particularly interesting. We predict a higher mass-loss rate for $\xi$ Boo A whereas \citet{Wood2010} suggest that it is $\xi$ Boo B that contributes most of the mass-loss in this system despite $\xi$ Boo A being extremely coronally active. The authors suggest that, above some activity level, mass-loss is inhibited by some mechanism.

In addition to comparing the mass-loss rates of individual stars, we can also compare the overall samples. Figure \ref{figWood} shows the mass-loss rates per unit surface area plotted in blue against X-ray flux for our models. Mass-loss rates are given by $\dot{M}=4\pi \rho(r_{\star}) v(r_{\star}) r_{\star}^2$ for the Parker model and taken directly from the CS model. X-ray fluxes are calculated by dividing the X-ray luminosities, from Eq. (\ref{eqXrayRatio}), by the stellar surface areas. Numerical values for the mass-loss rates can be found in table \ref{tabCalcVarsApp}. Additionally, we have overplotted the data presented by \citet{Wood2014} in red (see their Fig. 4). From their data, the authors suggest that winds from solar-type stars fall into two regimes separated by a so-called wind dividing line (dashed red line). For the stars below this line, they find a power law of $\dot{M} \propto F_X^{1.34\pm0.18}$ (solid red line). From our models, we find power laws of $\dot{M} \propto F_X^{0.30\pm0.02}$ and $\dot{M} \propto F_X^{0.67\pm0.10}$ for the Parker and CS models respectively when accounting for the low activity stars only (solid blue lines). If we include the whole sample in our fits, we obtain $\dot{M} \propto F_X^{0.38\pm0.01}$ and $\dot{M} \propto F_X^{0.79\pm0.05}$ respectively. We note that our quoted errors are in the fits only. Both our models predict increasing mass-loss rates with increasing X-ray activity. This general trend is in agreement with that of \citet{Wood2014} below their wind dividing line although the power law value they find is higher than ours. The main difference between our models and the results of \citet{Wood2014} is the behaviour at high X-ray flux. These authors suggest that solar-type stars can be divided into two wind regimes where the most active stars have lower mass-loss rates than some less active stars. The Parker model is unable to reproduce this behaviour above the wind dividing line unless the wind temperature is varied from star to star while the CS model shows some hints of this behaviour. It is clear that further astrospheric Ly$\alpha$ absorption observations of high activity stars are required to allow a more in depth study of winds from these stars.

It is also interesting to compare our results to those of \citet{Vidotto2013}, whose results we have plotted in Fig. \ref{figMPHZ} with square points. These magnetospheric sizes, for fictitious Earth-analogues in the HZs of M dwarfs, are calculated considering only the pressure contribution of the stellar magnetic field and neglect the wind ram pressure. These points are shown as upper limits since including the wind ram pressure would only decrease the size of the planetary magnetosphere. None of their planets are able to maintain an Earth-sized magnetosphere although several, at the higher host star mass end of the sample, are able to maintain a Paleoarchean-sized one. To make a comparison with the M dwarfs, we use the red curve, rather than the entire sample of solar-type stars. This is a fairer comparison since the solar-type stars span a large range of ages. When using the Parker model there may be a smooth transition between the solar and M dwarf samples. The situation is not as clear for the CS model. There appears to be a mass range, at around 0.6 $M_{\odot}$ - 0.8 $M_{\odot}$, where magnetospheric sizes peak. At lower masses, the HZ is too close to the star where the magnetic pressure can be high, especially at younger ages. At higher masses, the higher basal Alfv\'en wave flux is able to drive much stronger winds, as previously discussed. 0.6 $M_{\odot}$ - 0.8 $M_{\odot}$ therefore represents a stellar mass regime where neither the magnetic or ram pressures are too high.

\subsection{Evolution of planetary systems}
\label{subsecEvolution}
The analysis presented so far considers each system at a particular snapshot in time. However, the present conditions of a given system are determined by its history and will, in turn, determine the future evolution of the system. We already have a sense of how planetary magnetospheres around solar-type stars would evolve from the theoretical standoff curves plotted in Fig. \ref{figMPHZ}. In particular, we can look at the red, 8 Myr, curve to give us an idea of the level of magnetospheric protection available to young Earth-analogues. For the Parker model, magnetospheric sizes greater than 5$R_E$ are possible for all stellar masses above 0.6 $M_{\odot}$. However, the CS model predicts much reduced magnetosphere sizes for Earth-analogues around higher mass stars. Consequently, these planets may have a more difficult time retaining their atmospheres in the face of the harsher winds expected during the early, more active period of a star's life. However, our models predict that stars with a higher coronal activity present stronger winds. \citet{Wood2014} present evidence that young solar-type stars may have much weaker winds than expected. If this is indeed the case, young Earth-analogues may be more able to protect their atmospheres than initially thought.

It is also pertinent to consider the time evolution of the M dwarfs. If given enough time, could their magnetic activity decline to the point where the magnetic pressure is no longer the dominant pressure in the HZ? Additionally, could it decline to the point where an Earth-analogue in orbit could maintain a large magnetosphere? The former question is not easily answered. Stellar magnetic fields and winds are both linked to the stellar dynamo \citep{Schwadron2008} and both decline as a star ages along the main sequence \citep{Wood2005,Vidotto2014}. It is not clear when, or indeed if, the dominant pressure term might switch for a given spectral type. The latter question is easier to assess, at least in terms of magnetic pressure. \citet{West2008} studied how M dwarf activity lifetimes varied with spectral type. The authors find increasing activity lifetimes for later-type dwarfs; 8 Gyr for M7 dwarfs compared to 0.8 Gyr for M0. \citet{Vidotto2013} also studied the evolution of M dwarf magnetism using stellar rotation as a proxy for activity. The authors found that early- to mid-type M dwarfs require rotation periods of $\gtrsim$37-202 days in order for any Earth-analogues they host to maintain an Earth-sized magnetosphere. The required rotation period increases to $\gtrsim$63-263 days for late-type M dwarfs. Both results imply that only very old Earth-analogues around late type M dwarfs are able to maintain large magnetospheres.

Given the long timescales over which low mass stars remain magnetically active, it seems unlikely that an orbiting Earth-analogue could hold onto its atmosphere. Replenishment, via planetary volcanism, could be a solution to this problem \citep{Papuc2008,Jackson2008,Padhi2012}. However, this will only be effective if it is still occurring after stellar activity has declined sufficiently for the planet to maintain a large magnetosphere. This may be problematic given that volcanic activity has died down over geological timescales on Earth and Mars \citep{Werner2009}. Additionally, it is thought that water might be a necessary ingredient for plate tectonics \citep{Lieb2001}. If the atmosphere, and hence surface water, has already been lost, replenishment of the atmosphere by volcanism may be difficult. It is clear that planets in the HZs of M dwarfs would have greater difficulty maintaining their atmospheres. As such, planets around solar-type stars seem to be much more attractive targets for habitability in this regard.

\begin{figure}
	\begin{center}
	\includegraphics[width=\columnwidth]{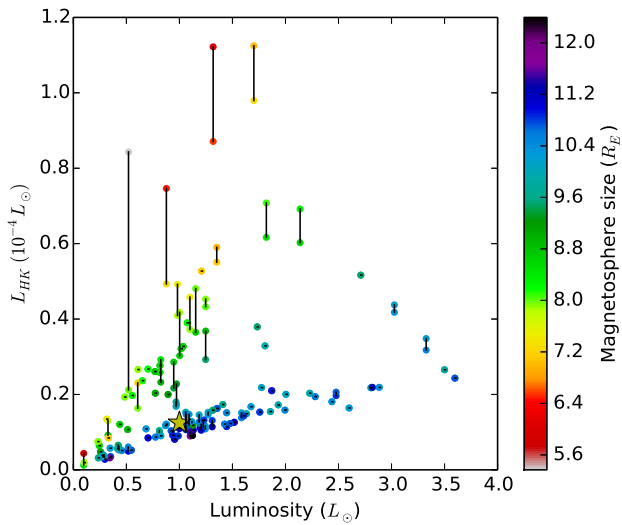}
	\end{center}
	\caption{Each star is plotted in Ca II H \& K luminosity-bolometric luminosity space and colour coded according to the magnetospheric size of its planet according to the Parker wind model. HD 107213 and HD 18256 have not been plotted because of their large luminosities. Low-luminosity and low-activity stars, i.e. the stars below the Vaughan-Preston gap, represent optimum observing targets.}
	\label{figParamSpace}
\end{figure}

\subsection{Observing Earth-analogues}
\label{subsecObservingHabPlanets}
Section \ref{subsecPRamHZ} demonstrates that, in general, exoplanets around solar-like stars do not simultaneously have both an Earth-like surface temperature and an Earth-sized magnetosphere. While most of the planets can maintain a Paleoarchean magnetosphere, it is still beneficial for them to have larger magnetospheres since the auroral opening, through which atmospheric leakage occurs, shrinks for larger magnetospheres \citep{Tarduno2010,Vidotto2013}. It would therefore be useful to identify the observational signatures of stars with weaker winds.

Figure \ref{figParamSpace} shows the parameter space that our sample occupies in terms of Ca II H \& K luminosity, $L_{HK}$, and bolometric luminosity with the points colour coded by magnetospheric size under the Parker model. The most striking feature is the presence of two distinct branches separated by the, so called, Vaughan-Preston gap \citep{Vaughan1980}. Many explanations have been proposed to explain the gap including different dynamos operating along each branch \citep{Bohm2007}, an abrupt change in chromospheric activity with stellar age \citep{Pace2009}, or multiple waves of star formation \citep{Mcquillan2013}.

In general, planets that orbit low chromospheric activity and low luminosity stars have greater magnetospheric protection which is evident from Fig. \ref{figParamSpace}. Planets orbiting stars on the lower, inactive, branch generally have larger magnetospheres. Whatever the cause for the Vaughan-Preston gap, it is clear that stars on the low activity branch are the optimum hosts of potentially habitable planets. The same trend can be seen when using the CS model although the data points have not been plotted for clarity.

\section{Conclusion}
\label{secConclusion}
We have presented a study on the effect of stellar winds on exoplanet magnetosphere sizes. This will better inform the search for extra-solar Earth analogues which may potentially be habitable. In particular, we have considered whether it is feasible for an exoplanet to simultaneously have an Earth-like surface temperature and Earth-sized magnetosphere. These constraints increase the chance of liquid water existing and help retain the planetary atmosphere respectively. Both are thought to be important to habitability. Since the exact size of magnetosphere required to protect a planetary atmosphere is not known, we consider two reasonable values of magnetospheric size. These are the present day magnetosphere size, 10.2 $R_E$, and its size 3.4 Gyr ago, thought to lie between 5 $R_E$ and 7 $R_E$.

Using the wind models of \citet{Parker1958} and \citet{Cranmer2011}, we estimated the magnetospheric extent of a hypothetical Earth orbiting each star in a sample of 124 solar-type stars. Within this modest sample, only a fraction of the planets were able to maintain a 10.2 $R_E$ magnetosphere. Most of them are able to maintain one of at least 7 $R_E$ within the Parker model although, within the CS model, a non-negligible number fall into, or below, the range of Paleoarchean magnetosphere sizes we consider. As stars age, the magnetospheric protection provided by an exoplanetary magnetic field of fixed strength will only increase thanks to the declining magnetic activity of the star. Our results suggest that a level of protection comparable to the early Earth's should be possible for planets orbiting stars of age greater than roughly 1.5 Gyr under the Cranmer \& Saar model and almost immediately after a star enters the main sequence according to the Parker model. 

The result is striking when compared to that of \citet{Vidotto2013}. Most terrestrial planet searches focus on M dwarfs because their low luminosities and masses are ideal for the transit and radial velocity techniques. However, from the point of view of atmospheric protection, young active M dwarfs can be less than ideal. Their habitable zones lie much closer in allowing the stellar magnetic pressure to compress planetary magnetospheres by a significant amount. Interestingly, when searching for well shielded planets in the habitable zone of stars, our work hints at the possibility of an optimum host star mass. For the Cranmer \& Saar model, planets orbiting 0.6 $M_{\odot}$ - 0.8 $M_{\odot}$ stars seem to have the largest magnetospheric sizes. Around this range of masses, the star is dim enough that the habitable zone lies far out, and hence stellar magnetic pressure is low, whilst having a low enough mass that convective jostling of flux tubes only drives a relatively weak wind. For the Parker model, magnetospheric sizes increase with increasing host star masses. Both models agree that, in general, solar analogues are more likely than M dwarfs to host planets with large magnetospheres and surface temperatures appropriate for liquid water. 

Other factors, such as the size of the auroral oval or whether the atmosphere is being replenished, will also determine if a stable atmosphere is present. These conditions vary from planet to planet and our chosen values of 5 $R_E$ to 7 $R_E$ and 10.2 $R_E$ should only be thought of as reasonable magnetospheric sizes on a sliding scale where larger sizes are clearly better. They should not be considered as strict criteria by which to judge magnetospheric protection. Our results indicate that planets around 0.6 $M_{\odot}$ - 0.8 $M_{\odot}$ stars on the low activity side of the Vaughan-Preston gap are the optimum observing targets for habitable Earth analogues. This, as well as the contrast between solar-type stars and M dwarfs, highlights how important it is to characterise the host star when considering habitability.

\begin{acknowledgements}
The authors are thankful for helpful comments from an anonymous referee. VS acknowledges the support of an STFC studentship. AAV acknowledges support from a Royal Astronomical Society Fellowship and an Ambizione Fellowship from the Swiss National Science Foundation. S.V.J. acknowledges research funding by the Deutsche Forschungsgemeinschaft (DFG) under grant SFB 963/1, project A16. Partly based on data from the Brazilian CFHT time allocation under the proposals 09.BB03, 11.AB05, PI: J.D. do Nascimento.
\end{acknowledgements}

\bibliography{StellarWindsPaper}
\bibliographystyle{aa}

\begin{appendix}
\section{Numerical Values}
\label{appTable}
We present the numerical results of this study for the full Bcool sample. 

\onllongtab{
\begin{longtable}{lccccccc}
\caption{Ram pressure, $P_{ram}$, exerted on a hypothetical Earth analogue located in the centre of the habitable zone, $r_{HZ}$, the corresponding magnetospheric size, $r_{MP}$, and the mass-loss rates, $\dot{M}$, for the Bcool sample of stars. Superscripts indicate the wind model used.}\\
\hline\hline
Star & 	$P_{ram}^{P}$ & $P_{ram}^{CS}$ & $r_{MP}^{P}$ & $r_{MP}^{CS}$ & $\dot{M}^P$ & $\dot{M}^{CS}$ & $r_{HZ}$\\
ID & [$P_{ram}^{E}$] & [$P_{ram}^{E}$] & [$R_E$] & [$R_E$] & [$\dot{M}_{\odot}$] & [$\dot{M}_{\odot}$] & [AU]\\
\hline
\endfirsthead
\caption{Continued from previous page.}\\
\hline
Star & 	$P_{ram}^{P}$ & $P_{ram}^{CS}$ & $r_{MP}^{P}$ & $r_{MP}^{CS}$ & $\dot{M}^P$ & $\dot{M}^C$ & $r_{HZ}$\\
ID & [$P_{ram}^{E}$] & [$P_{ram}^{E}$] & [$R_E$] & [$R_E$] & [$\dot{M}_{\odot}$] & [$\dot{M}_{\odot}$] & [AU]\\
\hline
\endhead
\hline
\endfoot
\hline
\endlastfoot
16 Cyg A	&	0.84/0.99	&	1.23/0.51	&	10.47/10.20	&	9.82/11.38	&	1.06/1.24	&	2.02/0.84	&	1.47		\\
16 Cyg B	&	0.88/1.00	&	0.29/0.37	&	10.38/10.18	&	12.50/11.99	&	0.88/0.99	&	0.37/0.48	&	1.3		\\
18 Sco	&	0.61/1.11	&	0.44/2.11	&	11.05/10.00	&	11.68/8.98	&	0.53/0.96	&	0.44/2.12	&	1.2		\\
5 Pegase	&	0.55/0.70	&	0.29/0.51	&	11.24/10.80	&	12.50/11.37	&	0.58/0.74	&	0.37/0.65	&	1.34		\\
EK Dra	&	7.28/14.91	&	75.36/96.42	&	7.31/6.48	&	4.95/4.75	&	5.28/10.81	&	59.38/75.98	&	1.09		\\
$\epsilon$ Eri	&	2.61/4.75	&	2.30/6.97	&	8.67/7.85	&	8.85/7.36	&	0.73/1.33	&	0.66/2.01	&	0.67		\\
HD 100180	&	0.72	&	0.98	&	10.74	&	10.21	&	0.79	&	1.25	&	1.36		\\
HD 10086	&	1.84	&	7.54	&	9.19	&	7.26	&	1.35	&	6.19	&	1.1		\\
HD 101501	&	4.35/7.91	&	6.27/17.88	&	7.96/7.21	&	7.49/6.29	&	2.18/3.96	&	3.65/10.41	&	0.91		\\
HD 103095	&	1.36	&	0.28	&	9.67	&	12.57	&	0.27	&	0.06	&	0.57		\\
HD 10476	&	1.00/1.32	&	0.43/0.74	&	10.18/9.72	&	11.69/10.70	&	0.36/0.47	&	0.17/0.30	&	0.76		\\
HD 10697	&	1.22	&	1.85	&	9.84	&	9.18	&	2.67	&	6.04	&	1.95		\\
HD 107213	&	0.44	&	2.56	&	11.65	&	8.7	&	1.87	&	14.97	&	2.73		\\
HD 107705	&	0.65	&	0.9	&	10.92	&	10.35	&	0.98	&	1.62	&	1.6		\\
HD 10780	&	2.33	&	2.66	&	8.83	&	8.64	&	0.99	&	1.27	&	0.83		\\
HD 111395	&	2	&	9.38	&	9.06	&	7	&	1.28	&	6.5	&	1.03		\\
HD 115404a	&	2.41	&	56.47	&	8.78	&	5.19	&	1.51	&	38.31	&	1.02		\\
HD 117936	&	2.91	&	1	&	8.51	&	10.17	&	0.62	&	0.22	&	0.58		\\
HD 120476a	&	8	&	0.67	&	7.19	&	10.87	&	2.27	&	0.22	&	0.67		\\
HD 122064	&	0.64	&	0.11	&	10.96	&	14.72	&	0.16	&	0.03	&	0.64		\\
HD 128165	&	1.85	&	0.57	&	9.18	&	11.18	&	0.41	&	0.13	&	0.59		\\
HD 12846	&	1	&	0.46	&	10.18	&	11.57	&	0.67	&	0.37	&	1.06		\\
HD 13043	&	1.07	&	0.75	&	10.06	&	10.68	&	1.69	&	1.56	&	1.65		\\
HD 131511	&	4.33	&	13.54	&	7.97	&	6.59	&	1.77	&	5.83	&	0.81		\\
HD 135101	&	0.66	&	1.31	&	10.9	&	9.73	&	0.76	&	1.93	&	1.4		\\
HD 13825	&	0.62	&	0.36	&	11.03	&	12.06	&	0.56	&	0.39	&	1.24		\\
HD 138573	&	0.79/0.85	&	0.43/0.53	&	10.58/10.44	&	11.71/11.31	&	0.68/0.73	&	0.45/0.55	&	1.2		\\
HD 145825	&	1.09/1.23	&	3.89/4.94	&	10.03/9.83	&	8.11/7.79	&	0.87/0.98	&	3.44/4.37	&	1.15		\\
HD 1461	&	0.61	&	0.37	&	11.03	&	12.01	&	0.59	&	0.43	&	1.28		\\
HD 149661	&	2.55	&	3.64	&	8.7	&	8.2	&	0.94	&	1.42	&	0.77		\\
HD 152391	&	3.56	&	13.1	&	8.23	&	6.63	&	1.66	&	6.5	&	0.87		\\
HD 15335	&	1.24	&	2.48	&	9.81	&	8.75	&	3.34	&	10.36	&	2.18		\\
HD 159909	&	1.41	&	1.35	&	9.6	&	9.68	&	1.59	&	1.94	&	1.39		\\
HD 160346	&	0.99	&	0.77	&	10.18	&	10.62	&	0.29	&	0.24	&	0.69		\\
HD 16141	&	0.85	&	1.63	&	10.45	&	9.38	&	1.33	&	3.48	&	1.65		\\
HD 16160	&	1.07	&	0.31	&	10.05	&	12.39	&	0.25	&	0.08	&	0.61		\\
HD 164595	&	0.68/0.74	&	0.38/0.46	&	10.84/10.70	&	11.97/11.57	&	0.59/0.63	&	0.39/0.47	&	1.2		\\
HD 166435	&	4.74/6.51	&	64.34/89.75	&	7.85/7.44	&	5.08/4.81	&	3.81/5.25	&	57.47/80.17	&	1.16		\\
HD 166620	&	0.73	&	0.16	&	10.73	&	13.84	&	0.22	&	0.05	&	0.69		\\
HD 166	&	4.5	&	29.73	&	7.92	&	5.78	&	2.3	&	16.03	&	0.91		\\
HD 171488	&	13.81/21.40	&	211.14/240.41	&	6.57/6.11	&	4.17/4.08	&	14.70/22.79	&	263.09/299.56	&	1.34		\\
HD 175726	&	4.07/4.40	&	53.66/62.81	&	8.05/7.95	&	5.24/5.10	&	4.11/4.45	&	62.45/73.10	&	1.3		\\
HD 179958	&	0.62/0.79	&	0.20/0.46	&	11.02/10.59	&	13.33/11.57	&	0.60/0.76	&	0.23/0.55	&	1.28		\\
HD 18256	&	0.78	&	18.18	&	10.61	&	6.27	&	3.95	&	116.89	&	3		\\
HD 1832	&	1.51	&	1.1	&	9.5	&	10.01	&	1.59	&	1.49	&	1.34		\\
HD 183658	&	0.60/0.76	&	0.36/0.61	&	11.08/10.65	&	12.08/11.05	&	0.59/0.75	&	0.42/0.72	&	1.29		\\
HD 185144	&	1.29/1.51	&	0.69/0.92	&	9.75/9.50	&	10.83/10.32	&	0.46/0.54	&	0.27/0.36	&	0.76		\\
HD 18803	&	1.03	&	1.36	&	10.13	&	9.67	&	0.73	&	1.12	&	1.09		\\
HD 190771	&	2.64/4.62	&	19.83/59.24	&	8.65/7.88	&	6.18/5.15	&	2.18/3.80	&	18.04/53.90	&	1.17		\\
HD 194012	&	1.18	&	8.24	&	9.9	&	7.16	&	1.71	&	13.88	&	1.57		\\
HD 196850	&	1	&	0.56	&	10.18	&	11.2	&	0.98	&	0.69	&	1.29		\\
HD 206860	&	2.64/4.26	&	32.00/83.96	&	8.65/7.99	&	5.71/4.86	&	2.49/4.02	&	33.41/87.66	&	1.25		\\
HD 208776	&	0.9	&	1.65	&	10.36	&	9.36	&	1.71	&	4.47	&	1.82		\\
HD 210277	&	0.55	&	0.23	&	11.23	&	13.03	&	0.43	&	0.21	&	1.14		\\
HD 213575	&	1.16	&	1.71	&	9.93	&	9.31	&	1.68	&	3.51	&	1.59		\\
HD 217107	&	0.31	&	0.2	&	12.39	&	13.25	&	0.28	&	0.21	&	1.24		\\
HD 217813	&	3.51	&	46.1	&	8.25	&	5.37	&	3.09	&	44.92	&	1.21		\\
HD 217877	&	0.93	&	2.2	&	10.3	&	8.92	&	1.24	&	3.75	&	1.51		\\
HD 218687	&	8.66/9.76	&	94.20/123.31	&	7.10/6.96	&	4.77/4.56	&	9.39/10.58	&	125.64/164.47	&	1.36		\\
HD 218868	&	0.79	&	0.61	&	10.57	&	11.03	&	0.5	&	0.45	&	1.03		\\
HD 219134	&	0.99	&	0.22	&	10.19	&	13.12	&	0.24	&	0.06	&	0.62		\\
HD 221146	&	0.78	&	2.18	&	10.6	&	8.93	&	1.4	&	5.22	&	1.76		\\
HD 221354	&	0.72	&	0.21	&	10.74	&	13.17	&	0.33	&	0.11	&	0.86		\\
HD 221356	&	0.85	&	0.62	&	10.46	&	11	&	0.98	&	0.9	&	1.4		\\
HD 221830	&	1.19	&	1.9	&	9.88	&	9.14	&	1.53	&	3.34	&	1.49		\\
HD 222143	&	2.69	&	25.07	&	8.63	&	5.95	&	2.29	&	23.45	&	1.19		\\
HD 225261	&	0.83	&	0.31	&	10.5	&	12.37	&	0.31	&	0.13	&	0.79		\\
HD 22879	&	1.85	&	0.47	&	9.18	&	11.53	&	1.64	&	0.58	&	1.24		\\
HD 24213	&	0.68/0.74	&	2.01/2.16	&	10.84/10.70	&	9.05/8.95	&	1.33/1.44	&	5.23/5.61	&	1.84		\\
HD 24496	&	0.61	&	0.66	&	11.03	&	10.91	&	0.36	&	0.42	&	0.98		\\
HD 25680	&	2.77	&	24.33	&	8.59	&	5.98	&	2.31	&	22.39	&	1.18		\\
HD 26965	&	1.24	&	0.57	&	9.82	&	11.19	&	0.43	&	0.22	&	0.75		\\
HD 28005	&	0.52	&	0.24	&	11.36	&	12.91	&	0.63	&	0.36	&	1.44		\\
HD 30562	&	0.59	&	1.85	&	11.11	&	9.18	&	1.31	&	5.4	&	1.96		\\
HD 30652	&	1.49	&	14.54	&	9.52	&	6.51	&	3.18	&	37.97	&	1.92		\\
HD 32147	&	1.02	&	0.29	&	10.14	&	12.5	&	0.25	&	0.07	&	0.62		\\
HD 34411	&	0.7	&	1.69	&	10.8	&	9.32	&	0.97	&	3.02	&	1.55		\\
HD 35296	&	3.18/4.03	&	84.50/137.43	&	8.39/8.06	&	4.86/4.48	&	4.66/5.92	&	142.14/231.19	&	1.57		\\
HD 3651	&	0.59/0.84	&	0.22/0.47	&	11.12/10.47	&	13.13/11.55	&	0.25/0.36	&	0.10/0.22	&	0.84		\\
HD 3765	&	0.45	&	0.12	&	11.62	&	14.48	&	0.13	&	0.04	&	0.68		\\
HD 377	&	8.5	&	81.52	&	7.12	&	4.89	&	8.33	&	95.02	&	1.28		\\
HD 3821	&	2.92	&	17.4	&	8.51	&	6.32	&	1.97	&	12.84	&	1.05		\\
HD 39587	&	4.50/6.45	&	33.61/70.96	&	7.92/7.46	&	5.66/5.00	&	4.03/5.76	&	34.80/73.46	&	1.22		\\
HD 45289	&	0.83	&	1.13	&	10.5	&	9.97	&	0.94	&	1.66	&	1.39		\\
HD 4614	&	0.69	&	0.53	&	10.82	&	11.29	&	0.68	&	0.63	&	1.29		\\
HD 4628	&	1.06/1.10	&	0.22/0.23	&	10.08/10.01	&	13.14/12.97	&	0.26/0.27	&	0.06/0.06	&	0.63		\\
HD 4915	&	0.91	&	1.93	&	10.33	&	9.11	&	0.52	&	1.2	&	0.97		\\
HD 5065	&	0.92	&	2.16	&	10.31	&	8.95	&	2.06	&	6.97	&	1.98		\\
HD 50692	&	0.95/1.02	&	0.75/0.93	&	10.27/10.13	&	10.68/10.29	&	0.97/1.05	&	0.93/1.16	&	1.32		\\
HD 56124	&	0.60/0.79	&	0.77/1.68	&	11.09/10.58	&	10.64/9.33	&	0.52/0.68	&	0.75/1.65	&	1.2		\\
HD 59747	&	6.24	&	8.04	&	7.5	&	7.19	&	1.69	&	2.27	&	0.66		\\
HD 71148	&	0.80/1.02	&	0.57/1.09	&	10.56/10.15	&	11.16/10.02	&	0.77/0.97	&	0.66/1.26	&	1.27		\\
HD 73344	&	1.51	&	14.66	&	9.5	&	6.5	&	2.11	&	23.5	&	1.54		\\
HD 73350	&	1.62/2.62	&	9.36/19.88	&	9.38/8.66	&	7.01/6.18	&	1.26/2.04	&	8.02/17.03	&	1.13		\\
HD 75332	&	2.63/3.34	&	47.38/93.79	&	8.66/8.32	&	5.35/4.77	&	4.49/5.70	&	95.54/189.14	&	1.71		\\
HD 76151	&	1.06/1.71	&	4.25/9.79	&	10.07/9.30	&	7.99/6.96	&	0.85/1.37	&	3.75/8.63	&	1.15		\\
HD 7727	&	1.06	&	1.39	&	10.07	&	9.62	&	1.5	&	2.46	&	1.55		\\
HD 78366	&	1.47/2.19	&	14.46/30.04	&	9.54/8.93	&	6.52/5.77	&	1.50/2.23	&	16.26/33.79	&	1.3		\\
HD 82106	&	4.48	&	1.68	&	7.92	&	9.33	&	0.9	&	0.35	&	0.56		\\
HD 8262	&	0.72	&	0.53	&	10.74	&	11.32	&	0.54	&	0.46	&	1.12		\\
HD 86728	&	0.69	&	0.29	&	10.83	&	12.49	&	0.55	&	0.28	&	1.16		\\
HD 88072	&	0.91/0.95	&	0.70/0.78	&	10.33/10.26	&	10.80/10.61	&	0.82/0.86	&	0.76/0.85	&	1.23		\\
HD 88230	&	4.52	&	0.06	&	7.91	&	16.32	&	0.41	&	0.01	&	0.37		\\
HD 88986	&	1.16	&	2.22	&	9.93	&	8.91	&	2.02	&	5.39	&	1.74		\\
HD 89269	&	1.24/1.35	&	0.79/0.97	&	9.81/9.68	&	10.58/10.22	&	0.82/0.89	&	0.63/0.77	&	1.05		\\
HD 9407	&	0.66	&	0.33	&	10.9	&	12.22	&	0.5	&	0.3	&	1.13		\\
HD 9562	&	0.72	&	1.95	&	10.74	&	9.1	&	2.02	&	7.86	&	2.21		\\
HD 98618	&	0.40/0.48	&	1.06/0.19	&	11.87/11.49	&	10.08/13.41	&	0.35/0.43	&	1.10/0.20	&	1.22		\\
HD 9986	&	0.63/0.90	&	0.61/1.61	&	10.99/10.35	&	11.04/9.39	&	0.56/0.80	&	0.63/1.65	&	1.22		\\
HIP 100970	&	1.11	&	1.94	&	9.99	&	9.11	&	1.69	&	4.05	&	1.62		\\
HIP 10339	&	3.78	&	18.59	&	8.15	&	6.25	&	2.05	&	10.89	&	0.94		\\
HIP 38228	&	3.54	&	23.85	&	8.24	&	6	&	2.08	&	14.9	&	0.98		\\
HIP 41844	&	0.72	&	1.4	&	10.74	&	9.62	&	0.85	&	2.12	&	1.42		\\
HIP 49350	&	0.88	&	1.49	&	10.38	&	9.52	&	0.64	&	1.21	&	1.09		\\
HIP 53721	&	0.7	&	1.63	&	10.79	&	9.38	&	0.9	&	2.64	&	1.48		\\
HIP 7244	&	2.88	&	20.76	&	8.53	&	6.14	&	1.95	&	15.12	&	1.05		\\
HIP 86974	&	0.93	&	1.24	&	10.3	&	9.81	&	1.88	&	3.79	&	1.88		\\
HR 1817	&	7.38/9.38	&	206.09/227.42	&	7.29/7.01	&	4.19/4.12	&	10.22/12.97	&	320.74/353.94	&	1.52		\\
$\kappa$ Cet	&	2.12/3.15	&	12.11/24.13	&	8.98/8.40	&	6.71/5.98	&	1.45/2.15	&	8.92/17.77	&	1.06		\\
$\tau$ Boo	&	0.92/0.99	&	3.56/4.51	&	10.32/10.19	&	8.23/7.92	&	2.19/2.37	&	10.48/13.26	&	2.03		\\
$\upsilon$ And	&	0.83/0.97	&	3.12/3.64	&	10.50/10.22	&	8.42/8.20	&	2.16/2.54	&	10.81/12.62	&	2.13		\\
$\xi$ Boo A	&	4.24/46.31	&	18.87/43.06	&	7.99/5.37	&	6.23/5.43	&	1.86/20.28	&	8.69/19.82	&	0.84		\\
$\xi$ Boo B	&	2.94/26.33	&	0.001/0.15	&	8.50/5.90	&	31.76/14.00	&	0.28/2.47	&	0.0001/0.01	&	0.36		\\
\label{tabCalcVarsApp}
\end{longtable}}

\end{appendix}
\end{document}